\documentclass{moriond}

\usepackage{float}
\def\be{\begin{equation}}
\def\ee{\end{equation}}
\def\bea{\begin{eqnarray}}
\def\eea{\end{eqnarray}}

\newcommand{\Photo}{}

\begin{document}
\vspace*{4cm}
\title{PHOTON TMD}

\author{H.~JUNG, S.~TAHERI~MONFARED, T.~WENING }

\address{Deutsches Elektronen-Synchrotron, D-22607 Hamburg,\\
II. Institut f\"ur Theoretische Physik, Universit\"at Hamburg}

\maketitle\abstracts{ We present a complete set of transverse momentum dependent and collinear parton  densities including the photon over a wide phase space. The photon density appears when evolving parton distributions with QED corrections. The photons are produced by perturbative radiation using the Parton Branching method. The QCD partons are constrained by a fit to HERA data. 
The lepton pair production at high masses is described within the extracted TMD densities.  }

\section{Motivation}
The high accuracy of current measurement requires
comparable precision in the phenomenological predictions. One of the limitations of the mainstream approach is the neglect of transverse degrees of freedom in the proton. The Parton Branching (PB) method \cite{Hautmann:2017xtx,Hautmann:2017fcj} 
aims
for correct treatment of the kinematics in collisions. This method was successfully applied to
describe the data from deep inelastic scattering (DIS) at HERA and Drell-Yan (DY) at the LHC
\cite{Martinez:2018jxt,Martinez:2019mwt,Martinez:2020fzs,Jung:2021mox} and can be further used for obtaining consistent predictions for a wide range of processes at the LHC, High Luminosity LHC (HL-LHC), High Energy LHC (HE-LHC) and future collider experiments. The advantage of applying transverse momentum dependent (TMD), instead of collinear parton density functions (PDFs) is that higher order corrections in form of parton showers are naturally included and that the uncertainty of parton showers is entirely tractable from the PB-TMDs.

The precise theoretical and phenomenological evaluations of the subleading electromagnetic processes are also needed to reach the same accuracy level of the experimental data. 
Including electroweak corrections to the pure QCD evolution is thus timely and is an important step in this direction.

\section{Collinear parton densities from PB method}
The first set of pure NLO QCD TMDs determined from a fit to precise HERA inclusive DIS measurements \cite{Abramowicz:2015mha} have been described in Ref. \cite{Martinez:2018jxt}. 
In this study, we focus on QED corrections using the PB method.

We apply charge-dependent LO QED splitting functions and the running QED coupling together with NLO QCD splitting kernels to the fit to HERAI+II data. The evolution is performed with Set~2 setting of Ref. \cite{Martinez:2018jxt}, a more detailed discussion is available in Ref. \cite{Jung:2021mox}. 
The fits of both pure QCD set \cite{Martinez:2018jxt} and QCD+QED set \cite{Jung:2021mox} to DIS measurements with the identical parametrization for all parton densities yield similarly good $\chi^2$-values. The resulting PB-TMD PDFs are available in the TMDlib package \cite{Abdulov:2021ivr}.

\begin{figure}[H]
\begin{center}
\begin{minipage}[c]{0.49\linewidth}
\includegraphics[width=0.95\linewidth]{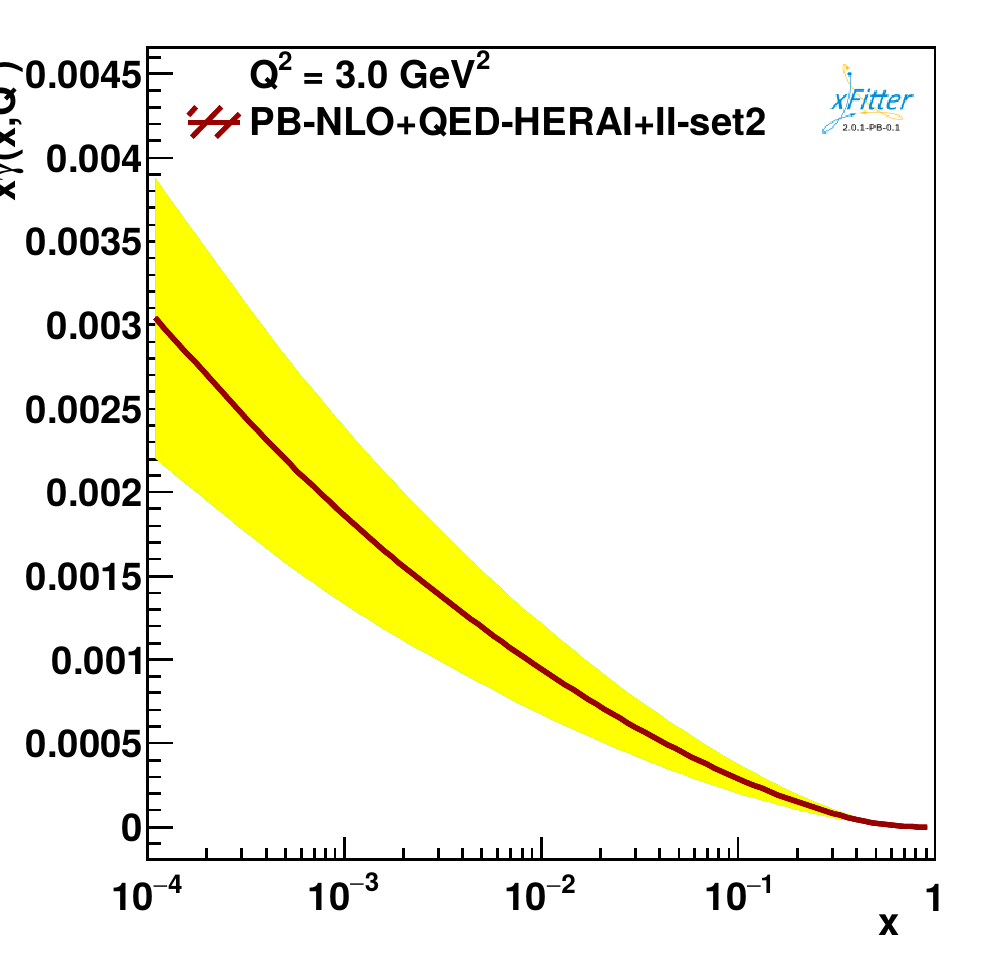}
\end{minipage}
\begin{minipage}[c]{0.49\linewidth}
\includegraphics[width=0.95\linewidth]{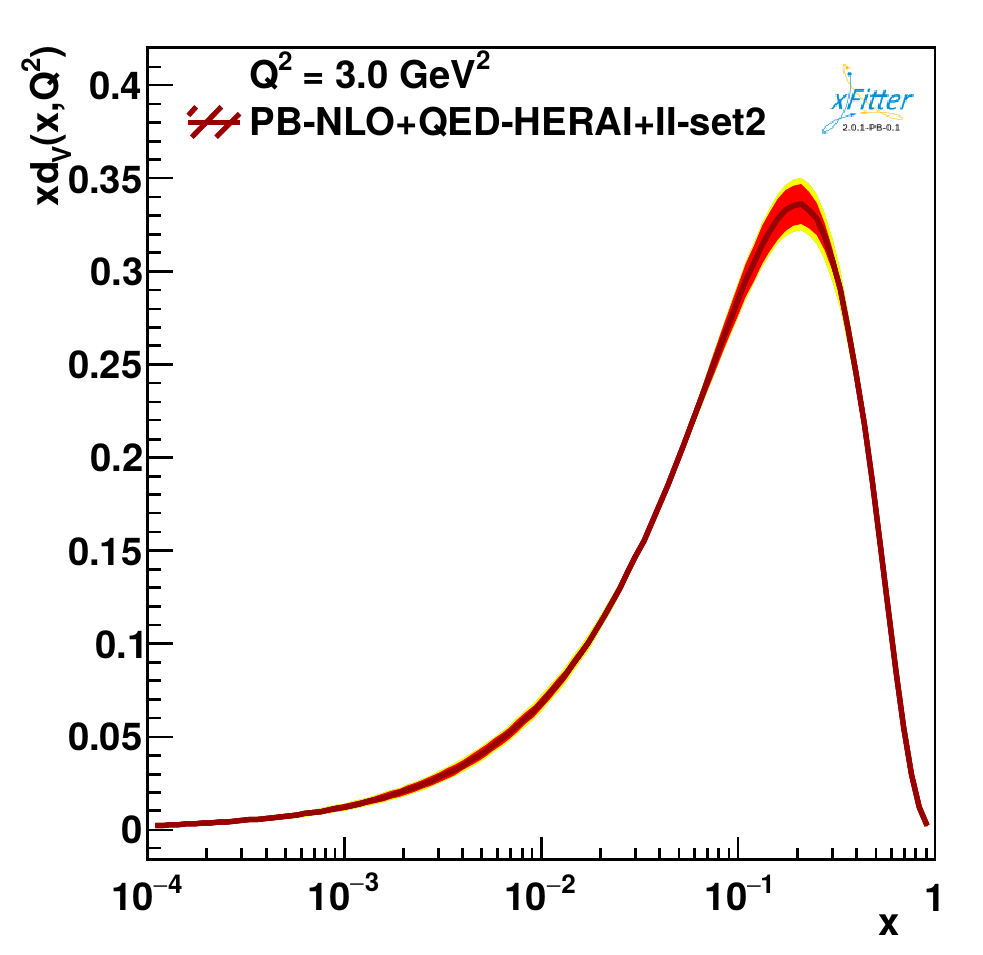}
\end{minipage}%
\caption{Collinear photon and valence down quark densities for $Q^2=3.0$ GeV$^2$. The red band shows the experimental uncertainty, the yellow band the model dependence.  }\label{fig:collinearPDF}
\end{center}
\end{figure}

In Fig. \ref{fig:collinearPDF} the photon and $d_v$ quark  densities are shown as functions of $x$ for $Q^2=3.0$ GeV$^2$, including the experimental uncertainties (red band) and the model uncertainties (yellow band).
The experimental uncertainties are obtained with the Hessian method (as implemented in the xFitter package \cite{Alekhin:2014irh}). The model dependence of the PDF fits is estimated by variation of  heavy flavor masses, the starting scale of the evolution and the $q_{cut}$ parameter. The parameter $q_{cut}$ is introduced in strong coupling definition to avoid the non-perturbative region while the argument in $\alpha_s$ is the transverse momentum.


 The small-$x$ photons originate mainly from radiation off large-$x$ valence quarks and thus the uncertainty of the small-$x$ photon density is rather large.

\section{TMDs from PB method}
Within the PB method both TMD parton densities as well as the collinear densities can be calculated.

\begin{figure}[H]
\begin{minipage}[c]{0.49\linewidth}
\includegraphics[width=\linewidth]{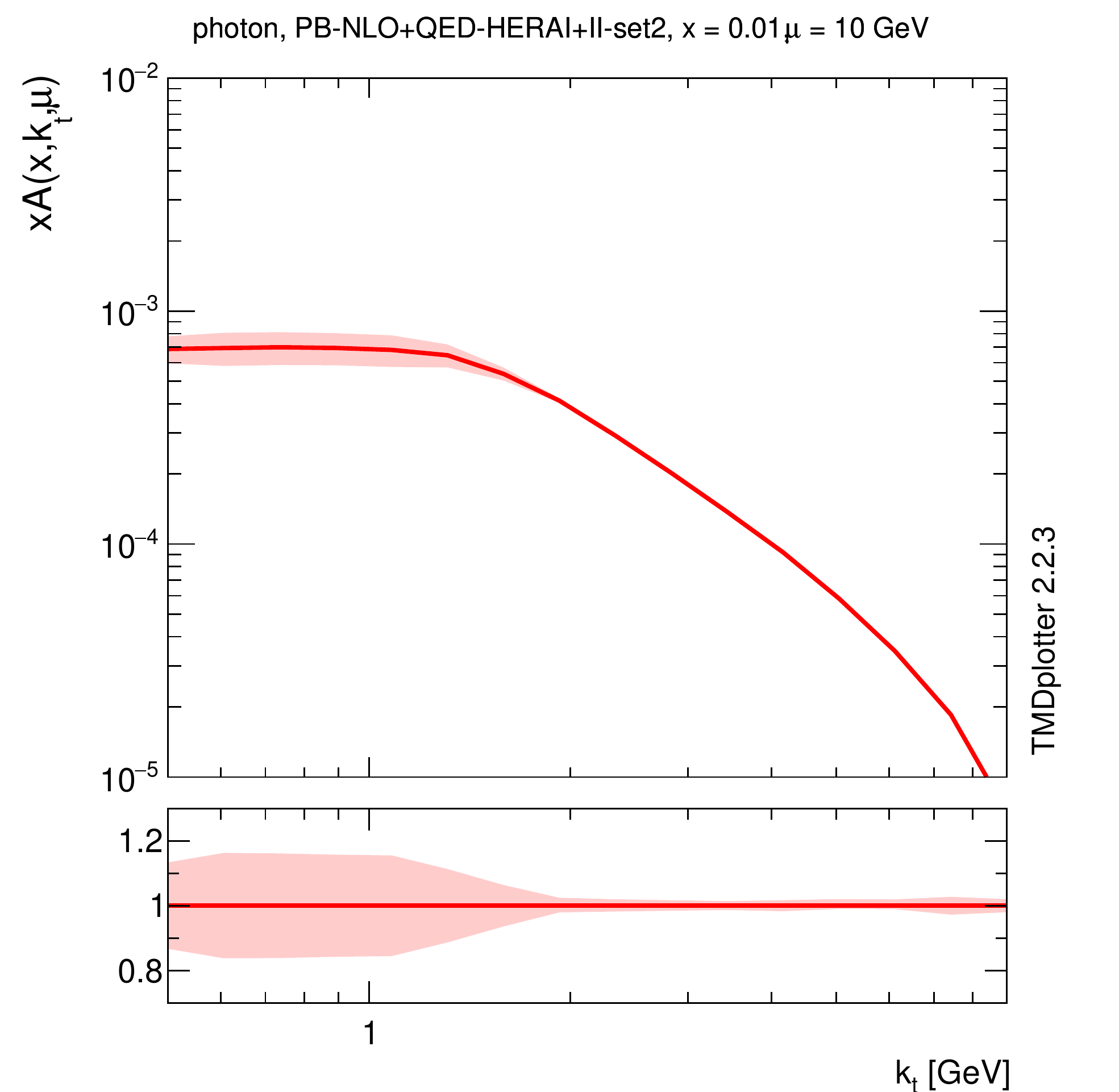}
\end{minipage}
\begin{minipage}[c]{0.49\linewidth}
\includegraphics[width=\linewidth]{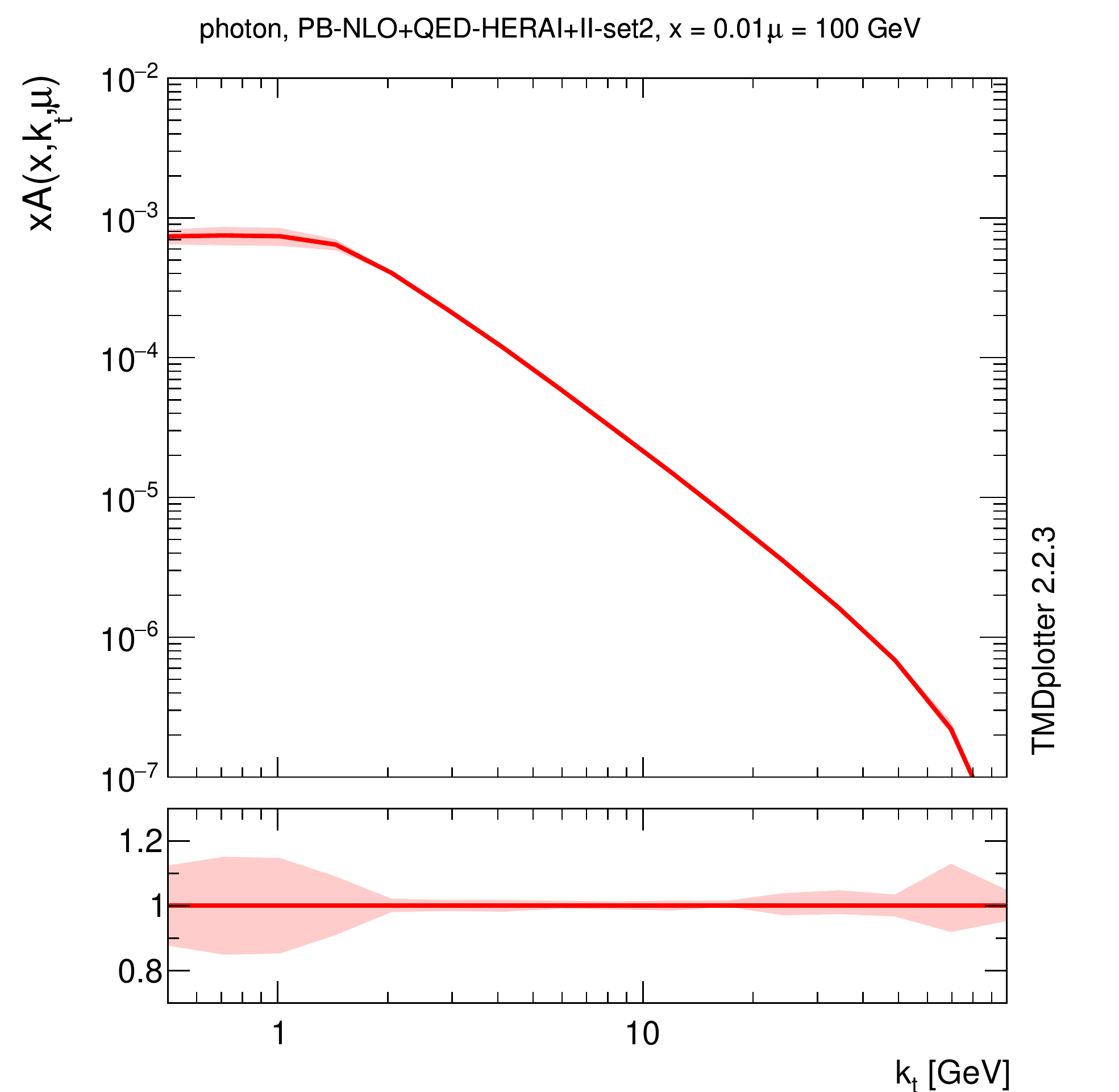}
\end{minipage}%
\caption{TMD photon density as a function of $k_t$ for $\mu=10$ and $100$ GeV. The total uncertainty is shown in the lower panels.}\label{fig:TMDPDF}
\end{figure}

In Fig. \ref{fig:TMDPDF}, the photon density together with the total uncertainty bands obtained from experimental and model uncertainties are shown as a function of the transverse momentum for $\mu=10,~100$ GeV. 
In lower panels of Fig. \ref{fig:TMDPDF}, we show a $k_t$-dependent uncertainty band  obtained from a  collinear parton density fit while splitting functions are also collinear. This happens because at different regions of transverse momentum, different $x$-values play a role in the evolution.

\section{Application to LHC measurements}
The CMS experiment \cite{Sirunyan:2018owv} has measured the production of muon pairs over a wide range of the dilepton invariant mass at $\sqrt{s}=13$ TeV. 
There is no measurement on the $p_t$-spectrum of high-mass DY pairs. This measurement can be proposed to constrain the photon TMD.

\begin{figure}[H]
\begin{center}
\includegraphics[width=0.45\linewidth]{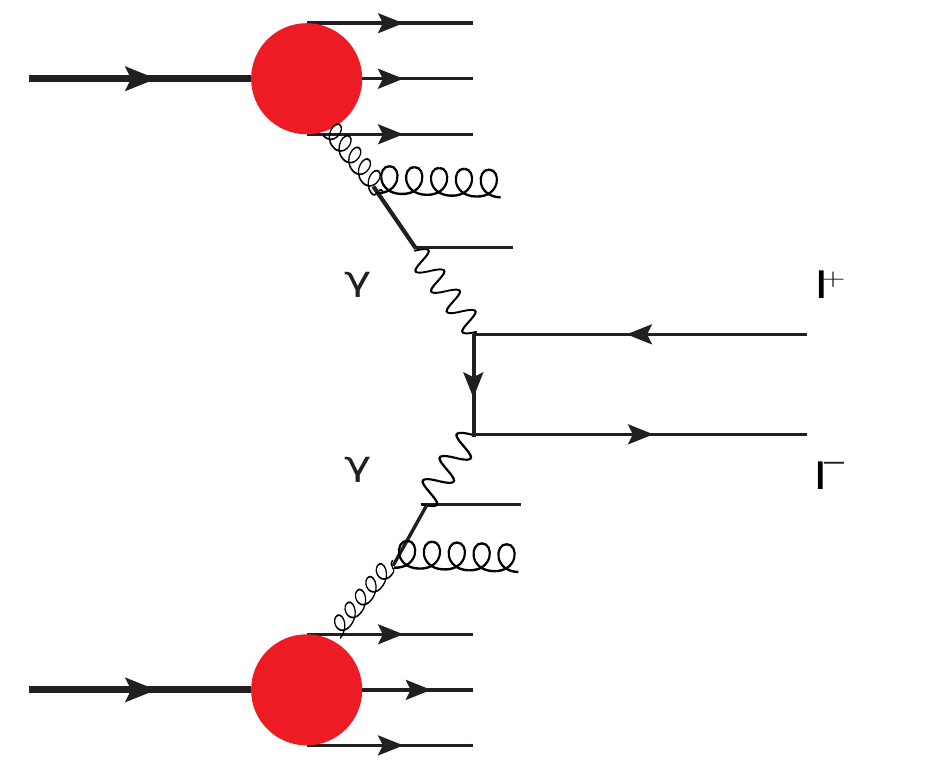}
\caption{Lepton pair production from dynamically generated photons}\label{fig:diagram-1}
\end{center}
\end{figure}

The contributions from photon-initiated (PI) channels to lepton pair production based on dynamically generated photons, as shown in  Fig.~\ref{fig:diagram-1}, are evaluated. 

\begin{figure}[H]
\begin{center}
\includegraphics[width=0.49\textwidth]{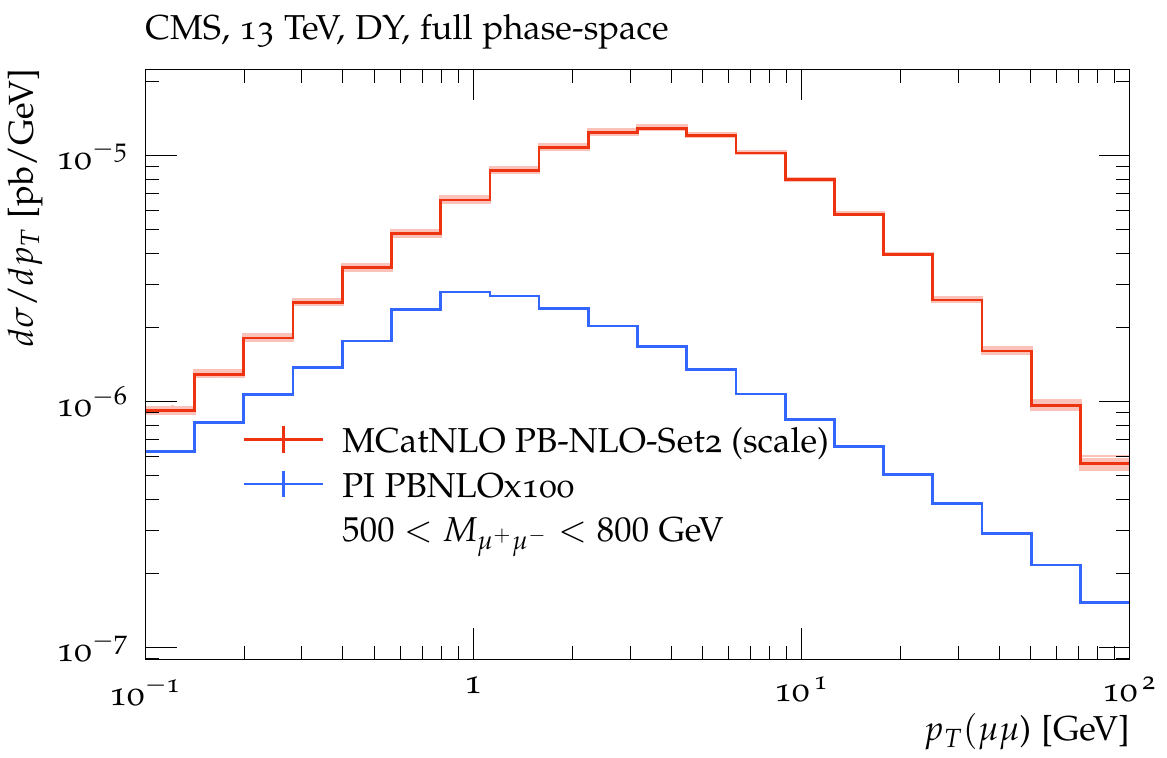}
\includegraphics[width=0.49\textwidth]{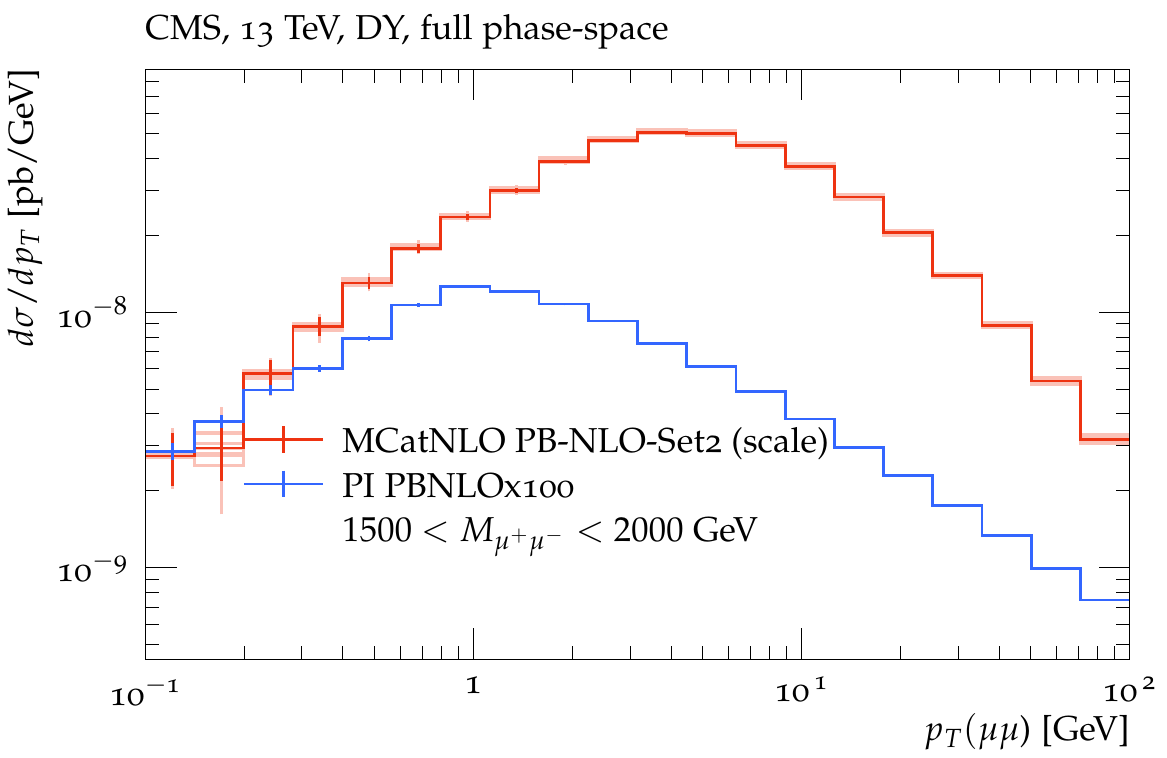}
\end{center}
\caption{Transverse momentum spectra of Drell-Yan and photon-initiated processes based on TMD PB-QED (Set2) at two different mass regions.}\label{fig:pt}
\end{figure}

In Fig. \ref{fig:pt} the transverse momentum spectra of the lepton pair in different high mass ranges of $500 < M_{\mu^+\mu^-} < 800$ and $1500 < M_{\mu^+\mu^-} < 2000$ are shown. We use NLO PB-TMD-QED and NLO matrix elements with the CASCADE3 MC event generator package \cite{Baranov:2021uol}. The DY ($qq\rightarrow l^+l^-$) contribution is shown in red and the scaled PI process ($\gamma \gamma\rightarrow l^+l^-$)  in blue. 
The transverse momentum spectra of the standard DY and PI leptons for very high DY mass values are similar only at low transverse momenta. The difference at high transverse momenta originates mainly from the hard matrix element process.

We present the first calculation of TMD photon density with the PB method. The outlook is a step towards calculation of the collinear and TMD heavy gauge boson parton densities which are essentially important for vector boson production at the LHC. 

\section*{Acknowledgments} 
This review is based on the work presented in Ref. \cite{Jung:2021mox}. 
We thank F. Hautmann for various discussions and comments on the manuscript.
STM thanks the Humboldt Foundation for the Georg Forster research fellowship.

\end{document}